\documentclass[mathleft,fleqn,%
]{an}
\usepackage{amsmath}
\usepackage{graphicx}
\usepackage[varg]{txfonts}
\usepackage{url}
\usepackage{natbib}
\bibpunct{(}{)}{;}{a}{}{,}
\begin{document} 
% The following seven commands are intended for editorial usage and
% should be ignored by the author(s).
\Pagespan{1}{}% Document's page range. 
% If second parameter is left empty, the last page is computed
% automatically.
\Yearpublication{2014}%
\Yearsubmission{2014}%
\Month{0}%   
\Volume{999}%  
\Issue{0}% 
\DOI{asna.201400000}% 

\title{Simultaneous spectra and radio properties of BL Lac's}
 \author{M.~Mingaliev
          \inst{1,2}
          \and
          Yu.~Sotnikova\inst{1}\thanks{sjv@sao.ru}
          \and
          T.~Mufakharov\inst{3}
         \and
         E.~Nieppola\inst{4,5}
	   \and
	   M.~Tornikoski\inst{4}
	   \and
	   J.~Tammi\inst{4,6}
	   \and
	   A.~L{\"a}hteenm{\"a}ki\inst{4,6}
	   \and
	   R.~Udovitskiy\inst{1}
	   \and
          A.~Erkenov\inst{1}
          }

\titlerunning{Simultaneous spectra and radio properties of blazars}
\authorrunning{M.\,Mingaliev}

   \institute{Special Astrophysical Observatory of RAS, Nizhnij Arkhyz, 369167 Russia\\
              \email{marat@sao.ru}
         \and
             Kazan Federal University, 18 Kremlyovskaya St., Kazan, 420008, Russia
          \and
             Shanghai Astronomical Observatory, Chinese Academy of Sciences, Shanghai 200030, China
          \and
             Aalto University Mets{\"a}hovi Radio Observatory, Mets{\"a}hovintie 114, 02540 Kylm{\"a}l{\"a}, Finland
             \and
             Finnish Centre of Astronomy with ESO (FINCA), University of Turku, V{\"a}is{\"a}l{\"a}ntie 20, FI-21500 Piikki{\"o}, Finland         
             \and
             Aalto University Department of Radio Science and Engineering, P.O. Box 13000, FI-00076 Aalto, Finland
             }

 
\received{XXXX}
\accepted{XXXX}
\publonline{in Astronomische Nachrichten, DOI 10.1002/asna.201713361}

\abstract{
We present the results of nine years of the blazar observing programme at the RATAN-600
 radio telescope (2005-2014). The data were obtained at six frequency bands (1.1, 2.3, 4.8, 7.7, 11.2, 21.7 GHz)
  for 290 blazars, mostly BL Lacs. In addition, we used data at 37 GHz obtained
quasi-simultaneously with the Metsahovi radio observatory for some sources.
The sample includes blazars of three types:
high-synchrotron peaked (HSP),
low-synchrotron peaked (LSP), and intermediate-synchrotron peaked (ISP).
We present several epochs of flux density measurements, simultaneous radio spectra, 
spectral indices and properties of their variability.
The analysis of the radio properties of different classes of blazars showed
that LSP and HSP BL Lac blazars are quite different objects on average. LSPs have
higher flux densities, flatter spectra and their
 variability increases as higher frequencies are considered.
On the other hand, HSPs are very faint in radio domain,
tend to have steep low frequency spectra, and they are less variable than LSPs at all frequencies.
Another result is spectral flattening above 7.7 GHz detected in  
HSPs, while an average LSP spectrum typically remains flat
at both the low and high frequency ranges we considered.
}

\maketitle

\section{Introduction}

Active galactic nuclei (AGN) are unique objects in the Universe, because they are compact and extremely luminous.
Their main observational properties can be explained by the presence of the super massive black hole at the centre
of the galaxies, surrounded by an accretion disk and by fast-moving clouds. They come in different flavors all 
of which can be explained in terms of the so-called ``unified scheme'' \citep{1995PASP..107..803U}, where the major 
difference between each of them is their orientation relative to the line of sight of the observer.

Blazars are the radio-loud subclass of AGN, characterized by strong non-thermal radiation across the entire 
electromagnetic spectrum \citep{1972ApJ...175L...7S,1994VA.....38...29K, 1998AJ....115.1253P}.
Strong variability at different timescales and wavebands is believed to be a result of relativistic motion of non-thermal plasma along the jet,
oriented at small angles to the observers line of sight \citep{1978bllo.conf..328B,1995PASP..107..803U}.
The average value of viewing angle of blazars is about $5^{\circ}$ according to the estimations made by \citealt{2013AJ....146..120L}.
The spectral energy distribution (SED) of blazars features two broad components:
the low-energy part (with the peak in the radio/infrared band) is believed to be formed by synchrotron emission 
and high-energy hump (with the peak in the gamma-rays) is usually explained in terms of inverse Compton radiation \citep[e.g.,][]{1994ApJ...421..153S,1996ApJ...463..444S} or hadronic processes \citep[e.g.,][]{2001APh....15..121M,2015MNRAS.448.2412P}.

Blazars are commonly distinguished as low-synchrotron peaked (LSP),
intermediate-synchrotron peaked (ISP), and high-synchrotron peaked (HSP),
depending on the peak frequency 
of the synchrotron component ($\nu_{peak}^{S}$) of their SED.
In this paper we adopt a blazar classification from the Fermi-LAT 3LAC catalogue \citep{2015ApJ...810...14A}, which uses  
the common convention by \citealt{2010ApJ...716...30A} to classify a blazar as an LSP for $\nu_{peak}^{S}<10^{14}~Hz$,
ISP for $10^{14}~Hz<\nu_{peak}^{S}<10^{15}~Hz$, and HSP for $\nu_{peak}^{S}>10^{15}~Hz$.

Blazars are historically divided in two main classes according to the presence or absence of the lines
in their optical spectra: BL Lacertae type objects (BL Lacs) have no lines or very weak ones 
(with the rest frame equivalent width EW$<$5 \AA),
while flat-spectrum radio quasars (FSRQs) exhibit normal quasar-like spectra 
with strong broad emission lines \citep{1991ApJ...374..431S,1995PASP..107..803U}.

Currently one of the most extensive list of blazars
is presented in the Roma-BZCAT catalogue by \citet{2009A&A...495..691M}.
This catalogue is based on
multi-frequency surveys and detailed checkout of the literature.
The total number of blazars currently
listed in the Roma-BZCAT (Edition 5.0) is more than 3500.
But only a relatively small number of objects have been
intensively observed at many frequencies simultaneously.
The spectral coverage of many of them is poor,
both in time and in frequency. 
Few observatories have long term monitoring programmes for blazars at radio wavelengths,
Mets{\"a}hovi Radio Observatory \citep{2004A&A...427..769T}, University of Michigan
Radio Observatory (UMRAO) \citep{1999ApJ...512..601A}, 
Owens Valley Radio Observatory (OVRO) \citep{2011ApJS..194...29R} 
and INAF-Istituto di Radioastronomia in Medicina and Noto \citep{2007A&A...464..175B} are among them.

The RATAN-600 radio telescope has been monitoring AGN on regular basis for more than 10 years.
In this paper we present observational results for 290 blazars obtained at six frequencies from 1.1 to 21.7 GHz with this telescope.
The main objective of our research is to obtain multi-frequency
information about radio properties of the blazar
subgroups by using a single instrument and simultaneous measurements to exclude
possible systematic errors as well as variability.
This unique data that we collected allowed us to study the spectral
properties for relatively large sample of blazars.

\section{Sample}
The sample consists of blazars - 
BL Lacs, BL Lac candidates, and blazars of an uncertain type -
with flux density more than 400 mJy (added to the sample in 2012) and more than 100 mJy (added in 2014) at 1.4 GHz,
selected from the Roma-BZCAT catalogue\footnote{http://www.asdc.asi.it/bzcat/}
\citep{2009A&A...495..691M} maintained by the ASI Science Data Center.
The first observations with the RATAN-600
started in early 2005 while the last sources were added to the
programme in early 2014. 
The sample is heterogeneous and contains sources both faint and extremely bright in the radio domain.
The median value of the flux density at 1.4 GHz is 0.3 Jy;
the minimum value is 0.002 Jy at this frequency (MS~0122.1$+$0903);
the maximum is 22.83 Jy (3C84).
The inclusion of some FSRQs in this sample can be explained by ambiguity
of blazar classification, as pointed out by \citet{1996MNRAS.281..425M} and \citet{2005MNRAS.356..225A},
and some BL Lacs have been classified as FSRQ or vice versa \citep{2012ApJ...748...49S}.
Originally (in 2005) we observed some FSRQs as the BL Lac sources or BL Lac candidates.
But in the blazars classification in the BZCATs later edition they were classified as FSRQs and
we have included them as such in order to stick to an uniform classification.
Since there are both blazars of uncertain type (27) and BL Lac candidates (12) in the sample,
it will not affect the analysis of the three subclasses
of blazars HSP, ISP and LSP.
Thus, the sample mainly consists of BL Lacs, BL Lac candidates,
blazars of unknown type, and additionally 14 FSRQs.

All objects of the sample were classified as low-synchrotron peak (LSP),
intermediate-synchrotron peak (ISP), and high-synchrotron peak (HSP) blazars: 
197 of them were detected by Fermi-LAT and classified according to the 3LAC.
For another 93 objects
we have used the ASDC SED builder\footnote{http://tools.asdc.asi.it/SED/} 
to fit the synchrotron part of the SED with the second or third degree polynom and get its maximum. 
The logarithm of the synchrotron peak frequency was calculated in the rest frame.
We did not classify two sources in our sample (PGC 59947 and BZB J1733+4519) 
because they do not have enough observational data points in their SED 
to extract the reliable $\log\nu_{peak}^{S}$ value.
The source list includes 149 LSPs, 62 ISPs, 77 HSPs and 2 blazars with unknown SED type.
The list of sources and their characteristics are presented
in Table~\ref{tab:param1}:

Col.~1~-- Source name,

Col.~2--3~-- R.A. and Dec (J2000.0),

Col.~4~-- Number of observing epochs at RATAN,

Col.~5~-- Redshift $z$,

Col.~6~-- Logarithm of the synchrotron peak frequency $\nu_{peak}^{S}$ in Hz,

Col.~7~-- SED class from 3LAC, an asterisk was used to denote blazars, for which the values of
$\log\nu_{peak}^{S}$  were calculated using ASDC SED builder,

Col.~8~-- AGN class from BZCAT (Edition 5.0) \cite{2015Ap&SS.357...75M}.

\begin{figure}
\centerline{\includegraphics[width=90mm]{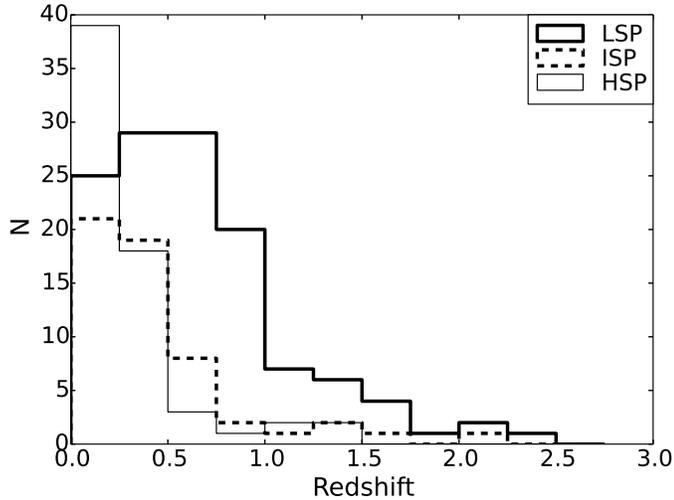}} 
\caption{Redshift distribution for different blazar subclasses of the sample.}
\label{fig:red}
\end{figure}

The redshifts for objects were mostly obtained from the
Roma-BZCAT, and for some sources from Simbad\footnote{http://simbad.u-strasbg.fr/simbad/}.
Redshifts are known for 238 sources, which is a relatively large number, considering
the difficulty of obtaining this information for BL Lacs.
The redshift distribution is presented in Fig.~\ref{fig:red}.
The redshift values range from $z$=$0.018$ to $2.277$,
with the mean equal to ~0.45 and median to ~0.36. Redshifts have not been measured for 52 objects;
13 of them are HSP objects, 8 are ISPs and 31 are LSPs.

\section{Observations}

The observations were carried out with the RATAN-600 radio telescope
during 2005--2014. Systematic monitoring was carried out during the period
2006-2008 for roughly one third of the sources.
Observations were done in the transit (meridian) mode with the north and south
sectors of the antenna \citep{1979S&T....57..324K,1993IAPM...35....7P}.
An angular resolution in this mode of observations depends on a declination
a source being observed. The FWHM in right ascension (RA) is given in Table~\ref{tab:radiometers};
an angular resolution in declination is three to five times worse than in RA.
Sources were observed from 3 to 15 times in each epoch to increase the reliability
of results and because of the weather or the receiver conditions.

Observation were carried out at six frequencies:
4.8, 7.7, 11.2, and 21.7 GHz (cryogenically cooled radiometers),
and 1.1 and 2.3 GHz (uncooled radiometers). The 4.8 GHz cooled radiometer
is a noise added radiometer (NAR), while other radiometers are designed
according to the beam-switching scheme.
All the radiometers were designed as the ``direct amplification Dicke type''
receivers. We use the data acquisition and controling
system for all continuum radiometers, as described by \citet{2011AstBu..66..109T}.

The experimental data were processed with using the modules of the FADPS
(Flexible Astronomical Data Processing System) standard reduction
package by \citet{1997ASPC..125...46V};
this is a reduction system for data from the broadband continuum
radiometers of the RATAN-600 secondary mirror. The processing
methods are described in \citet{2007ARep...51..343M,2012A&A...544A..25M}.
The following eight flux density calibrators were used to calculate
the calibration coefficients in the scale by \citet{1977A&A....61...99B}: 3C48,
3C138, 3C147, 3C161, 3C286, 3C295, 3C309.1, and NGC 7027.
In addition, we used the traditional RATAN-600 flux density
calibrators at low elevations: J0240$-$23, J1154$-$35, J0521$+$16,
and the source J0410$+$76 at high elevations.
Measurements of calibrators were corrected for angular size and linear polarization
(where necessary), following the data from \citet{1994A&A...284..331O} and \citet{1980A&AS...39..379T}.

The detection limit for RATAN-600 single sector
is approximately 8~mJy under good conditions
at 4.8 GHz and at an average antenna elevation ($\delta$$\sim$$40^{\circ}$). 
At other frequencies the detection limit
is presented in the Table~\ref{tab:radiometers}.
This value depends on the atmospheric extinction and the effective
area on the antenna elevation $H$ (from $10^{\circ}$ up to $90^{\circ}$ above the horizon)
at the corresponding frequencies.

The 37 GHz observations were made with the 13.7 m diameter Aalto
University Mets\"ahovi radio telescope, which is a radome enclosed
Cassegrain type antenna situated in Finland (60 d 13' 04'' N, 24 d 23'
35'' E). The measurements were made with a 1 GHz-band dual beam
receiver centered at 36.8 GHz. The HEMPT (high electron mobility
pseudomorphic transistor) front end operates at room temperature. The
observations are Dicke switched ON--ON observations, alternating the
source and the sky in each feed horn. A typical integration time to
obtain one flux density data point is between 1200 and 1600 s. The
detection limit of our telescope at 37 GHz of the order of 0.2 Jy
under optimal conditions. Data points with a signal-to-noise ratio $<$ 4
are handled as non-detections.
The flux density scale is set by observations of the HII region DR
21. Sources NGC 7027, 3C 274 and 3C 84 are used as secondary
calibrators. A detailed description of the data reduction and analysis
is given in \citet{1998A&AS..132..305T}. The error estimate in
the flux density includes the contribution from the measurement rms
and the uncertainty of the absolute calibration.

%The 37 GHz observations were obtained with the 13.7 m diameter Mets\"ahovi radio telescope.
%The measurements were made with a 1 GHz-band dual beam receiver centred at 36.8 GHz.
%The flux density scale is set by observations of the H II region DR21, 
%with a known flux density of 17.9 Jy at 37 GHz \citep{1998A&AS..132..305T}. 
%Sources NGC 7027, 3C 274 and 3C 84 are used as secondary calibrators. 
%A detailed description of the data reduction and analysis is given in \citet{1998A&AS..132..305T}.

\begin{table}
\caption{\label{tab:radiometers}RATAN-600 continuum radiometers.
Where $f$$_0$ -- central frequency, $\Delta$$f_0$ -- bandwidth,
$\Delta$$F$ -- flux density detection limit per beam,
and BW -- beam width -- an angular resolution in RA (an angular resolution in declination is three to five times worse than in RA).}
\centering
\begin{tabular}{rlcr}
\hline
 $f_{0}$ & $\Delta$$f_{0}$ & $\Delta$$F$ &  BW \\
  GHz    &   GHz           &  mJy/beam   &  arcsec \\
\hline
 $21.7$ & $2.5$  &  $70$ & 11 \\
 $11.2$ & $1.4$  &  $20$ & 16 \\
 $7.7$  & $1.0$  &  $25$ & 22 \\
 $4.8$  & $0.9$  &  $8$  & 36 \\
 $2.3$  & $0.4$  &  $30$ & 80 \\
  $1.1$  & $0.12$ &  $160$& 170 \\
\hline
\end{tabular}
\end{table}

\section{Results}

\subsection{Flux densities}
\label{flux}

The flux densities and the instantaneous spectra at several epochs of the sample sources (except the 37 GHz) are published in the BL Lac
database\footnote{www.sao.ru/blcat/} maintained by the Special
Astrophysical Observatory. The database is constantly updated
with more data which is freely available and is
described by \citet{2014A&A...572A..59M}.

Almost all sources have complete data
at frequencies 4.8 and 7.7~GHz. When available, we also added near-simultaneous
37 GHz data from the Aalto University Mets\"ahovi Radio Observatory database.
The data were considered near-simultaneous if they were taken within two weeks
of the RATAN observation.
The occasional absence of data at certain frequencies is a result of data exclusion because of the
partial resolution of a source at some frequencies,
a source too weak to be measured reliably, a strong influence of
man-made interferences at 1.1 and 2.3 GHz, or due to 
strong interference from geostationary satellites at 11.2~GHz
(between $-10^{\circ}$ and $0^{\circ}$ degrees of declination).
The values of the standard error of
fluxes for the most sources are: 5-20 \% for 11.2, 7.7,
and 4.8 GHz, 10-35 \% for 2.3, 1.0, and 21.7 GHz.

\begin{figure}
\resizebox{\hsize}{!}{\includegraphics[width=68mm]{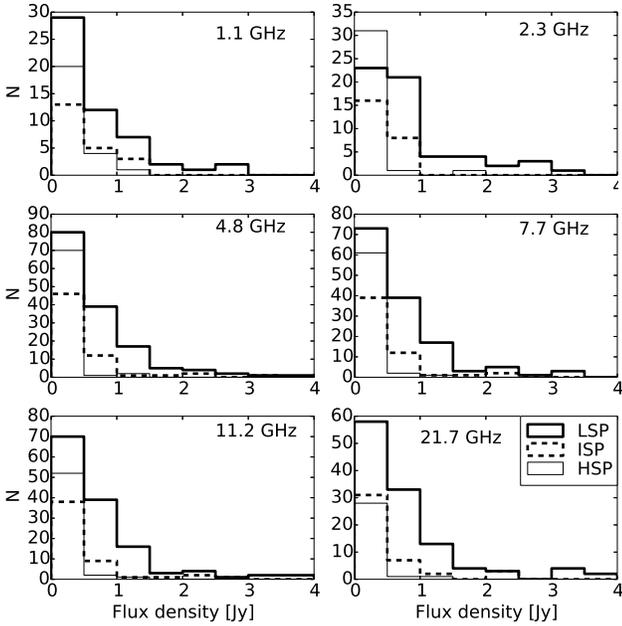}} 
\caption{The average flux density distribution at the RATAN frequencies by SED subclass
 (some bright sources with the average flux density above 4 Jy are excluded from the plot for the clarity of display).}
\label{fluxhist}
\end{figure}

The distributions of the average flux densities by source
class at all RATAN frequencies are shown in Fig.~\ref{fluxhist}. We used a bin size of 0.5
with the middle value of the bin shown in the graph. The average flux densities have been
calculated for $S/N>4$ detections only. The sample sources are typically extremely faint,
especially HSPs. Only one HSP (Mrk 501) in the sample has an average flux density
reaching 1 Jy at most frequencies.
This fact underlines the exceptional nature of this sample. Sources
this faint are rarely targeted in observational campaigns, and never
before have had their radio spectra determined this extensively.

The brightest source in the sample by far is 3C 84 ($S_{aver}$=23.18 Jy at 7.7 GHz,
clearly more than the second brightest source BL Lac with 6.66 Jy).
At 1.1 GHz 1828+487 is the brightest, but likely only because 3C 84
does not have data at that frequency (1.1 GHz observations were
interrupted in 2012). The flux density of 1828+487 quickly decreases
for frequencies higher than 1.1 GHz due to its steep spectrum. The medians of the average
fluxes by frequency and class are listed in Table~\ref{tab:ave}. The average
flux density values are not drawn from the same distribution for the
different SED classes according to the Kruskal-Wallis test. Pairwise comparisons confirm that LSPs are significantly
brighter than HSPs and ISPs at all frequencies ($P<0.001$).

We included the medians of the flux density values at 37 GHz in Table~\ref{tab:ave}.
These data come from the brightest sources at this frequency,
for example, only for 4 HSP objects and 10 ISP.
This can obviously bias the interpretation,
so the flux at 37 GHz looks high in general.

\begin{table*}
\caption{\label{tab:ave} The medians of the flux density values for the SED classes for each frequency,
$N$ is the number of sources used to estimate the medians that were calculated
for $S/N>4$ detections only (the brightest source $3C84$ was excluded from calculation).}
\centering
\begin{tabular}{lllllllllllllll}
\hline\hline
Class & $S_{1.1}$ & N & $S_{2.3}$ & N & $S_{4.8}$ & N & $S_{7.7}$ & N & $S_{11.2}$ & N & $S_{21.7}$ & N & $S_{37}$ & N \\
 & [Jy] & & [Jy] & & [Jy] & & [Jy] & & [Jy] & & [Jy] & & [Jy] & \\
\hline
 HSP  & 0.27 & 25 & 0.09 & 33 & 0.06 & 73  & 0.06 & 64  & 0.06 & 55  & 0.16 & 30  & 0.64 & 4 \\
 ISP  & 0.53 & 22 & 0.42 & 25 & 0.35 & 63  & 0.31 & 56  & 0.30 & 52  & 0.38 & 43  & 0.63 & 10 \\
 LSP  & 0.54 & 56 & 0.72 & 61 & 0.68 & 153 & 0.73 & 146 & 0.68 & 141 & 0.71 & 121 & 1.37 & 43 \\
\hline
\end{tabular}
\end{table*}

%________________________________________________________$

\subsection{Spectral indices}

We calculated spectral indices for the frequency intervals $1.1$-$7.7$
and $7.7$-$21.7$~GHz to investigate the spectral behavior of blazars at
low and high frequencies. 
We used spectral indices measured in the 1.1-7.7~GHz and in 7.7-21.7~GHz frequency intervals as the base range.
We also take the indices in the 2.3-7.7~GHz and in 7.7-11.2~GHz range as low and high frequency spectral indices for some sources 
that did not have measurements in the base range to increase the significanse of the statistics.
The spectral index $\alpha$ ($S \propto \nu^{\alpha}$) was calculated by:
\begin{equation}
\label{1}
\alpha=\frac{\log S_{2}-\log S_{1}}{\log{\nu}_{2}-\log{\nu}_{1}},
\end{equation}
where $S{_1}$ is the flux density at the frequency $\nu{_1}$, and $S{_2}$  the flux
density at the frequency $\nu{_2}$. Indices were calculated for $S/N>4$ detections only.

We present results for \textit{average} and \textit{instantaneous} low and high frequency spectral indices.

\begin{figure}
\centerline{\includegraphics[width=95mm]{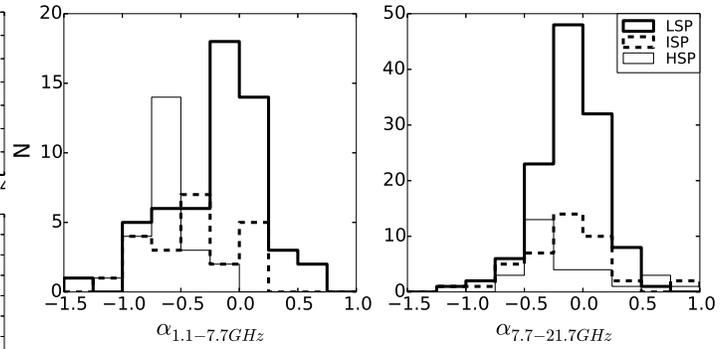}} 
\caption{Distribution of the \textit{average} low and high frequency spectral indices by SED subclass.}
\label{alphas}
\end{figure}

The \textit{average} spectral indices are listed in Table~\ref{tab:index}.
Fig.~\ref{alphas} shows the distribution of the average spectral index across the sample.
We have found that the spectrum of all three types of blazars we considered flattens
at the higher frequencies. The spectral indices are closer to zero and have less scatter in 7.7-21.7~GHz and
11.2-37~GHz intervals in comparison to index measured in 1.1-7.7~GHz frequency
range. Some care is required in interpreting this result, because of the data at 37~GHz
and relatively high variability indices at 21.7 and 37~GHz (see Table~\ref{tab:vary} in the next subsection).

\begin{table}
\caption{\label{tab:index}Average and median spectral indices by classes for 1.1-7.7, 7.7-21.7
and 7.7-37 GHz frequency intervals. N - is the number of the \textit{instantaneous} measurements used for calculations.}
\centering
\begin{tabular}{llcc}
\hline
Class & N       & $\alpha$ & $\alpha$ \\
        &          &  average   &  median  \\
\hline
$1.1-7.7$~GHz \\
\hline\hline
 HSP      & 55  & -0.61 & -0.55 \\
 ISP      & 65  & -0.28 & -0.26 \\
 LSP      & 190 & -0.05 & -0.03 \\
\hline
$7.7-21.7$~GHz \\
\hline
 HSP      & 85  & -0.27 & -0.30 \\
 ISP      & 154 & -0.19 & -0.16 \\
 LSP      & 668 & -0.12 & -0.10 \\
\hline
$7.7-37$~GHz \\
\hline
 HSP      & 25  & -0.15 & -0.14  \\
 ISP      & 29  &  0.03 & -0.01 \\
 LSP      & 265 &  0.01 &  0.02 \\
\hline
\end{tabular}
\end{table}

\begin{figure}
\resizebox{\hsize}{!}{\includegraphics{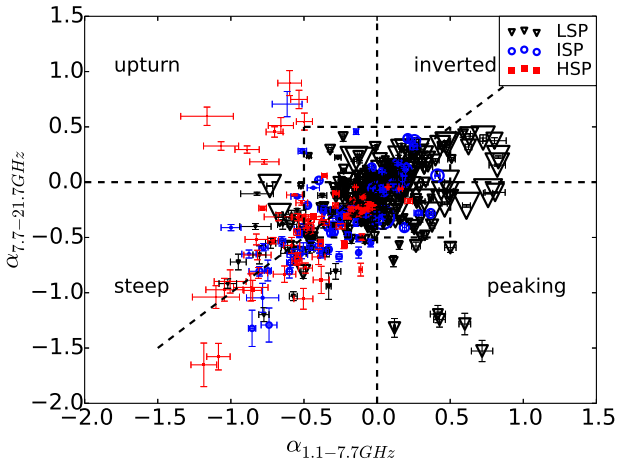}}
\caption{The ``radio colour'' plot for \textit{instantaneous} low and high frequency spectral indices. Black reverse triangles denote LSPs, blue circles - ISPs and red squares for HSPs. The size of the marks is related to the flux density at 7.7 GHz.}
\label{spec_corr}
\end{figure}

Now we will describe results on \textit{instantaneous} low and high frequency 
spectral indices for blazars in our sample. The \textit{instantaneous} spectral index term
 refers to a spectral index obtained at one epoch, 
thus the number of the \textit{instantaneous} spectral indices is equal to the number of epochs of observations for each source.
We plotted the \textit{instantaneous} low and high frequency spectral indices in a
two-colour diagram (or ``radio colour'') in Fig.~\ref{spec_corr}. This kind of representation
 has been previously 
used by, e.g., \citet{2006MNRAS.371..898S} and \citet{2008MNRAS.386.1729T}.
It is a convenient way to describe the shape of the radio spectrum. The
dashed lines divide the diagram into four quadrants signifying a steep,
upturning, inverted or peaking spectra as marked in the figure. 
The dashed inclined line is an one-to-one correspondence line, which
is an indicator of the spectral behavior.
The position of points relative to this line reflects the
character of the spectral slope change (steeper or flatter).
The box in the center of the plot denotes an area for the flat spectra
 (with $|{\alpha_{radio}|}<0.5$), it is usually referred as the ``blazar box''.
Most of the spectra corresponding to LSPs (82 \%) and ISPs (63 \%) are flat, while
only 41 \% of HSPs have a flat spectrum. 
For 44 \% of HSPs and 34 \% of the ISPs 
the low frequency spectrum is, in fact,
steep. Table~\ref{tab:spec_corr} lists the
number of objects in each segment of the plot: inside the
blazar box (flat spectrum), and in each quadrant outside
the blazar box.

For 35 out of 63 HSPs the spectrum turns flatter at 7.7 GHz,
i.e., they are situated above the one-to-one correspondence
line in Fig.~\ref{spec_corr}. 
Nine HSP sources even have a clear upturn.
The behaviour of ISPs and LSPs is different: for more than a half of LSPs (183/250) 
and ISPs (48/70) the spectrum turns steeper at 7.7 GHz.
HSPs show more often spectral
flattening beyond 7.7 GHz (or even inverted), while the LSP spectra get steeper at
that point.
LSPs and HSPs clearly occupy different regions of the two-colour plot,
with LSPs confined to the right edge and reaching high values of
spectral indices (${\alpha_{radio}}>-0.5$),
while HSPs generally fall into ``steep'' and ``upturn'' quadrant.

Kruskal-Wallis pairwise comparisons confirm that the distributions 
of the distances from the one-to-one line 
are not significantly different either for HSPs and ISPs or ISPs and LSPs,
but instead are significantly different for HSPs and LSPs ($P=0.05$).

The Pearson product-moment correlation coefficient between the \textit{instantaneous} low and high frequency
spectral indices is relatively high and significant only for ISPs ($r=0.60$).
The Spearman rank correlation coefficient values are: 0.28 for HSPs, 0.63 for ISPs and 0.37 for LSPs;
and significant (at the 0.05 level) only for ISPs and LSPs. 
%Also, if we remove the outlier LSP 0108+388 in the far lower right corner 
%and outlier ISP 0148+14 in the upper \textcolor{red}{left} corner in the Fig.~\ref{spec_corr} from the sample,
%the Pearson and Spearman correlation coefficients becomes more significant: $r=0.38$
%($P=0.008$), $\rho$=$0.40$ ($P=0.005$) for LSPs and $r=0.71$
%($P=0.001$), $\rho$=$0.75$ ($P=0.0005$) for ISPs. 
The correlation means that the typical radio spectrum
of ISPs and LSPs can be better approximated with a single power-law, while
such a model does not describe HSPs spectra accurately.

From both Fig.~\ref{alphas}
and Fig.~\ref{spec_corr} it is clear that the three SED classes have different
distributions of the low and high frequency spectral indices, that fact is 
confirmed by the Kruskall-Wallis test.
%which shows that three subclasses come from different populations for both low and high spectral indices.
%and populations are not significantly different in case of high spectral index (${\alpha_{7.7-21.7}}$ and ${\alpha_{11.2-37}}$).
The difference is also evident
in the \textit{average} spectral indices in Table~\ref{tab:index}.

\begin{table}
\caption{\label{tab:spec_corr}Blazars distribution in the ``radio colour'' plot in Fig.~\ref{spec_corr} by type of the radio spectrum: flat, inverted, peaking, steep or upturn. ``Flat'' radio spectrum refers to objects inside the blazar box, with $|{\alpha_{radio}|}<0.5$. The number of measured \textit{instantaneous} spectral indices for each blazar subclass is presented, the fraction of measurements belonging to the specific spectrum type of is given in brackets.}
\centering
\begin{tabular}{lccc}
\hline\hline
 Spectral  shape & \multicolumn{3}{c}{Number of indices}\\
 & LSP  & ISP & HSP \\
\hline
flat         &  206 (82\%) & 44 (63\%) & 26 (41\%)\\
inverted  &  7   (3\%)    & 0 (0\%)      & 0 (0\%) \\
peaking   &  12  (5\%)   & 0 (0\%)      & 0 (0\%) \\
steep      & 25  (10\%) & 24  (34\%) & 28  (44\%) \\
upturn     &  0   (0\%)    & 2 (3\%)     & 9 (14\%) \\
\hline
%all       & \textbf{383} &  \textbf{250} &  \textbf{70} &  \textbf{63}   \\
%\hline
\end{tabular}
\end{table}

%________________________________________________________

\subsection{Variability}

In order to characterize the variability properties of the sources
at the various frequencies we have computed the variability and modulation indices.
The first one takes into account measurement uncertainties,
while the modulation index is less sensitive to outliers.
The variability index was calculated as \citep{1992ApJ...399...16A}:

\begin{equation}
\label{2}
V_{S}=\frac{(S_{max}-\sigma_{S_{max}})-(S_{min}+\sigma_{S_{min}})}
{(S_{max}-\sigma_{S_{max}})+(S_{min}+\sigma_{S_{min}})}
\end{equation}
where $S_{max}$ and $S_{min}$ are the maximum and minimum value of the flux density, respectively,
at all epochs of observations; $\sigma_{S_{max}}$ and $\sigma_{S_{min}}$
are their errors. This prevents one from overestimating
$V_{S}$ when there are observations with large uncertainties in the dataset.
The negative value of $V_{S}$ corresponds to the case
where the flux error is greater than the observed scatter in the data.

The modulation index, defined as the standard deviation of the flux density divided by the mean flux density, was calculated as in \citealt{2003A&A...401..161K}:

\begin{equation}
\label{3}
M=\frac{\sqrt{\frac{1}{N}\sum_{i=1}^N(S_i-\frac{1}{N}\sum_{i=1}^N{S_i})^2}}{\frac{1}{N}\sum_{i=1}^N{S_i}}
\end{equation}

The medians of the variability and modulation indices for the SED classes by frequency are listed in Table~\ref{tab:vary}.
Almost 50 \% of HSP objects, 48 \% of ISPs, and 56 \% of
LSPs show variability $M\ge0.2$.
At high frequencies (7.7, 11.2 and 21.7~GHz) only a small number of each type of blazar
show variability above 0.7-0.8.
The most variable object at all frequencies except 1.1 GHz is
AO 0235+164 (27 epochs at high frequencies).
It has a bright peaking spectrum at two epochs due to a passing flare
in addition to the generally flat spectrum.
A large number of observations
increases the probability of finding it in active states, such as the two
peaking spectra contributing to the high variability index.
However,
at 1.1 GHz the source has little variability. The most variable
source at 1.1 GHz is 2E 0323+0214 with $M_{1.1}$=0.54.

\begin{table*}
\caption{\label{tab:vary} The medians of the variability $V_{S}$ and modulation indices $M$ for three blazar subclasses.
N is the number of epochs used to estimate the medians.}
\centering
\scriptsize
\begin{tabular}{rrrrrrrrrrrrrrrrrrrrrr}
\hline\hline
	   && \multicolumn{2}{c}{1.1 GHz} && \multicolumn{2}{c}{2.3 GHz} && \multicolumn{2}{c}{4.8 GHz} && \multicolumn{2}{c}{7.7 GHz} && \multicolumn{2}{c}{11.2 GHz}  & &  \multicolumn{2}{c}{21.7 GHz} && \multicolumn{2}{c}{37 GHz} \\
	   &&            &   &           &   &           &   &           &   &            &   &            &   &          &  & & & & & & \\
Class && N & $M$ && N & $M$ && N & $M$ && N & $M$ && N & $M$ && N & $M$ && N & $M$ \\
\hline
HSP && 30 & 0.15 &&  34 & 0.14 & &  49 & 0.11 & &  47 &  0.12 && 43  & 0.13 && 32  & 0.19 &&  3 & 0.20 \\
ISP && 24 & 0.16 &&  24 & 0.14  & & 39 & 0.11 & & 39 &  0.12 && 38   & 0.12 && 31  & 0.18 &&  5 & 0.27 \\
LSP && 48 & 0.17 &&  61 & 0.14 & &114  & 0.12 & &113 & 0.13 && 110   & 0.14 && 101 & 0.21 && 30 & 0.21 \\
	   &&            &   &           &   &           &   &           &   &            &   &            &   &          &  & & & & && \\
\hline\hline
      && N & $V_S$ & & N & $V_S$ & & N & $V_S$ & & N & $V_S$ & & N & $V_S$ & & N & $V_S$ & & N & $V_S$ \\
\hline
HSP && 30 & 0.15 & &  34 & 0.10 & & 49 & 0.10 & &  47 & 0.14 & & 43  & 0.11 & & 32  & 0.13 & &  3 & 0.10 \\
ISP && 24 & 0.15 & &  24 & 0.10 & & 39 & 0.11 & & 39 & 0.13 & & 38  & 0.17 & & 31  & 0.17 & &  5 & 0.19 \\
LSP && 48 & 0.17 & &  61 & 0.17 & &114 & 0.14 & &113 & 0.17 & & 110 & 0.17 & & 101 & 0.20 & & 30 & 0.29 \\
\hline
\end{tabular}
\normalsize
\end{table*}

\begin{figure}
\resizebox{\hsize}{!}{\includegraphics[width=68mm]{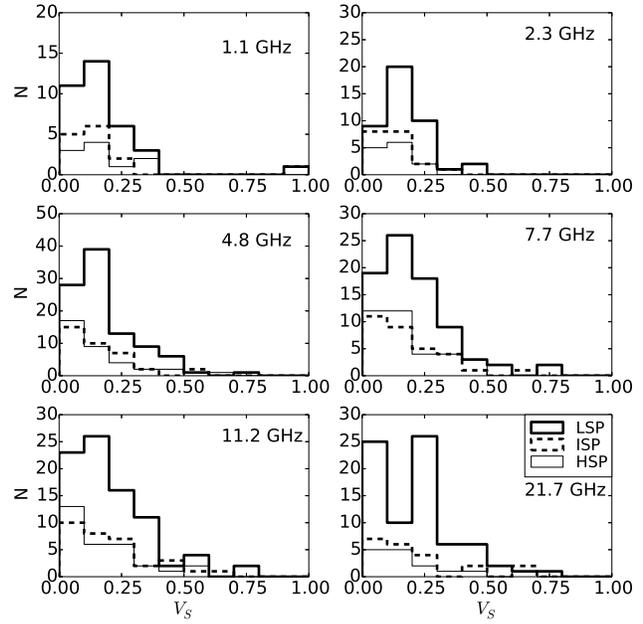}}
\caption{The variability index distributions at the RATAN frequencies by SED subclass.}
\label{varhist}
\end{figure}

\begin{figure}
\resizebox{\hsize}{!}{\includegraphics[width=68mm]{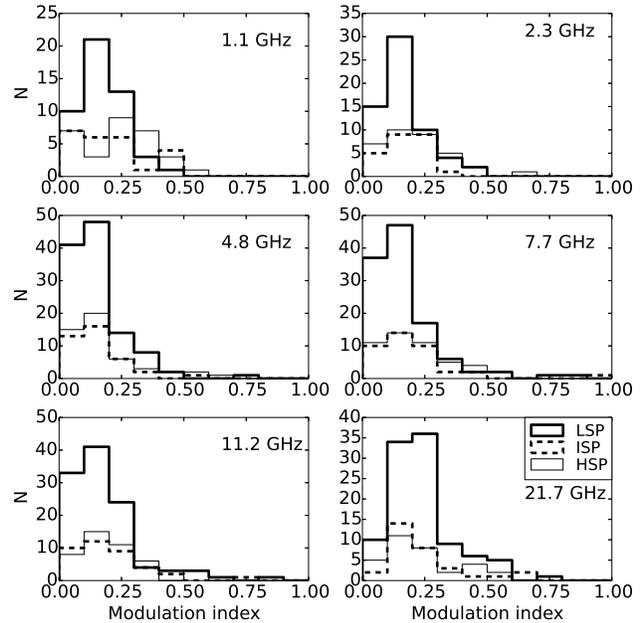}}
\caption{The modulation index distributions at the RATAN frequencies by SED subclass.}
\label{modhist}
\end{figure}

The HSPs and ISPs have lower values of the modulation index at six frequencies
than LSPs.
On the whole, blazars of all three types are less variable at 4.8 GHz,
according to the median values of both the modulation and the variability indices.
The modulation indices are not drawn from the same distribution only at low frequencies
(1.1 and 2.3 GHz) for the LSP and HSP classes according to the Kruskal-Wallis test.

%\textcolor{red}{When using the variability index it should be remembered that it is depends
%on the number of datapoints \citep[e.g.,][]{2007AJ....133.1947N,2014MNRAS.438.3058R},
%and depends on the range of time scales or activity states of the source.
%As we stated in Sect.~\ref{flux}, LSPs are significantly brighter than ISPs and HSPs,
%which means that they tend to get observed more often (Table~\ref{tab:nvar}).
%For example, at 4.8 GHz, the mean number of
%observational epochs for LSPs is 6, while for HSPs only 3.
%There is a correlation between $N_{obs}$--$V_{S}$,
%according to the Spearman rank correlation test: the correlation
%coefficient is 0.2 ($P=0.001$).
%If the object is often observed, it will be often detected in the different states.
%We are not able to examine strong variability that usually detected
%above the synchrotron peak frequency \citep{2014MNRAS.438.3058R,2014MNRAS.439..690H},
%where the shortest wavelength component becomes more optically thin, because even highest
%observation frequency of the RATAN is two orders of
%magnitude less than synchrotron peak frequency even for LSPs.}

\section{Discussion}
We find interesting differences between the blazar classes in the radio
colour plot (Fig.~\ref{spec_corr}). Almost half of HSPs (28 out of 63) in the sample are found
to be steep spectrum sources. Only the minority of LSPs and ISPs
(25/250 and 24/70, respectively) fall into ``steep'' quadrant,
the most of them have a flat radio spectrum.
The latter is not surprising, because a flat radio spectrum was a
standard selection criterion of the early radio-selected BL Lac
samples. Most of those sources are today classified as LSPs.

%The spectral shapes of the two classes seem to be inherently
%different. HSPs more often display a spectral flattening (or even invert) after
%7.7 GHz, whereas the spectra almost of half ISPs and LSPs steepen at the same point.
%However, we acknowledge that this result is not very robust due to
%the low flux levels of the HSP sample. A faint HSP with a spectral
%steepening at 7.7 GHz would be unlikely to be detected at 21.7 GHz.
%This possibility is strengthened by the weak linear correlation between
%the flux density at 7.7 GHz and the distance to the one-to-one line in Fig.~\ref{spec_corr},
%which depicts the intensity of the steepening or flattening.
%The Spearman correlation between these quantities gives $\rho=-0.3$ both for HSP and ISP
%($P=0.05$ and 0.01, respectively).
%This result is in good agreement with the \cite{2008MNRAS.386.1729T},
%where the linear correlation coefficient between the flux density at 33 GHz
%and spectral index $\alpha^{33}_{15}-\alpha^{4.8}_{1.4}$ is equal to 0.027.

The results obtained in this study indicate that the steep
and flattening spectra of HSPs is real. One possibility is
that there is an extended component in HSP BL Lacs
producing optically thin synchrotron radiation, while the flat 
radio spectrum of LSP BL Lacs can be explained by the compact nuclear core emission.
\citet{2012MNRAS.422.2274C}
investigated the dependence between the core/extended
morphology and radio spectral indices using 20 GHz data
from the Australia Telescope Compact Array for the AT20G
sample. They conclude that $\alpha =-0.46$ is an effective
division point between the compact and extended sources. They also
note that a considerable fraction of the extended sources
(i.e., $\alpha <-0.46$) show spectral flattening in the
higher frequencies. These sources are presumed to be
composed of a compact core with an extended component,
such as a steep-spectrum quiet jet, lobe or a hot spot.
At high radio frequencies this steep spectrum component
weakens while the compact, flat spectrum component
strengthens, and we observe this as spectral flattening.
This situation could happen if the jet is not aligned with the 
observers line of sight, but the viewing angle is still not very big
(close to $30^{\circ}$ in case of BL Lacs (\cite{1993ApJ...407...65G, 1999MNRAS.304..160J}).
That explains the flattening of HSPs in our sample.

%\textcolor{red}{The different distribution for the variability criteria
%($V_{S}$ and $M$) at low frequencies for HSPs and LSPs also can hint at differences in the
%mechanisms that produce the radio spectra of them.}

The largest variability in synchrotron radiation is observed
after the peak. Assuming that the flat radio spectrum is a result
of superimposed individual synchrotron components, the variability
should, on average, increase toward higher frequencies. That is 
observed in all three types of blazars in our sample at 4.8-21.7 GHz (for LSPs even until 37 GHz) (see Table~\ref{tab:vary}).
Secondly, according to our data, both variability and modulation indices for HSPs
at all seven frequencies (except the modulation index at 2.3 GHz) are less than for LSPs. 
That result is also expected, because sources with flat spectrum 
are more variable than sources with steep spectrum, 
this result is also in good agreement with \cite{2006MNRAS.370.1556B, 2006MNRAS.371..898S, 2008MNRAS.386.1729T}.

\section{Summary}

We have presented the results of an observational campaign on BL Lac blazars carried
out between 2005 and 2014 with the RATAN-600 radio telescope.
The flux densities at frequencies 1.1, 2.3, 4.8, 7.7, 11.2 and 21.7 GHz were
measured at several epochs simultaneously. This data set gives us the
opportunity to investigate the differences in the radio spectra of all
BL Lac types, including the radio-faint HSPs that are rarely targeted
by radio studies. We have made two main conclusions:

\begin{enumerate}
\item LSP and HSP BL Lac blazars are quite different objects on average. LSPs have
higher flux densities (the median of the flux density value is $S_{7.7}$=0.73 Jy),
flatter spectra and their
 variability increases towards higher frequencies.
HSPs are very faint in radio domain (the median is $S_{7.7}$=0.06 Jy),
tend to have steep low frequency spectra, and they are less variable than LSPs at all frequencies. 
\item 
LSP blazars have flat radio spectra at both low and high frequency ranges that we considered,
while the steep low frequency spectra of HSPs (and ISPs) turns flatter above 7.7 GHz. 
But despite the considerable decrease in the spectral index (``flattening'') 
at higher frequencies in HSPs, 
we note that the spectrum of LSPs is still flatter than HSPs.
\end{enumerate}

\acknowledgements
The RATAN-600 observations were carried out with the financial support of
the Ministry of Education and Science of the Russian Federation (14.518.11.7054)
and Russian Foundation for Basic Research (12-02-31649).
The  authors (MGM, TVM) acknowledge support through the Russian
Government Program of Competitive Growth of Kazan Federal University.

\bibliographystyle{an} % bst-файл, задающий стиль оформления библиографии
\bibliography{mingaliev} %\bibliographystyle{mn2e-long}

\appendix
 
\onecolumn   %%%% ESSENTIAL, these!!!!
\setcounter{table}{5}
\begin{table}
\caption{\label{tab:param1}Blazar sample and some parameters.}
\centering
\begin{tabular}{llllllll}\hline
Source & R.A. (J2000) & Dec (J2000) & N & z & $\log$ $\nu_{peak}^{S}$ & SED class & AGN class\\
\hline
%\endfirsthead
%\caption[]{continued.}\\
%\hline
%\hline
%Source & R.A. (J2000) & Dec (J2000) & N & z & $\log$ $\nu_{peak}^{S}$ & SED class & AGN class\\
%\hline
%\endhead
%\hline
%\endfoot
NRAO 5                  & 00:06:13.9 & -06:23:35  &  15  &  0.347   & 13.16 & LSP*&   BL Lac        \\
MS 0011.7+0837          & 00:14:19.7 & +08:54:01  &  6   &  0.162   & 12.44 & LSP*&   BL.gal.domin   \\
PKS 0017+200            & 00:19:37.8 & +20:21:45  &  3   &  0       & 12.32 & LSP &   BL Lac         \\
PKS 0019+058            & 00:22:32.5 & +06:08:05  &  16  &  0       & 13.11 & LSP &   BL Lac          \\
2MASX J00323309-2849200 & 00:32:33.0 & -28:49:20  &  1   &  0.324   & 14.70 & ISP &   BL Lac           \\
RXS J0325.2+1515        & 00:35:14.9 & +15:15:04  &  8   &  1.280   & 15.10 & HSP &   BL Lac             \\
1ES 0037+405            & 00:40:13.7 & +40:50:04  &  1   &  0       & 13.16 & LSP*&   bl.un               \\
RXS J0045.3+2127        & 00:45:19.2 & +21:27:42  &  1   &  0       & 16.00 & HSP &   BL Lac            \\
PKS 0047+023            & 00:49:43.3 & +02:37:04  &  11  &  0       & 13.63 & LSP &   BL Lac              \\
PKS 0048-097            & 00:50:41.2 & -09:29:06  &  14  &  0.200   & 14.61 & ISP &   BL Lac              \\
NPMIG -09.0033          & 00:56:20.0 & -09:36:29  &  3   &  0.103   & 15.55 & HSP &   BL.gal.domin     \\
RXS J0058.2+1723        & 00:58:16.7 & +17:23:13  &  2   &  0       & 16.28 & HSP*&   BL Lac           \\
B2 0103+33              & 01:06:00.2 & +34:02:03  &  1   &  0.579   & 13.52 & LSP*&   bl.un            \\
Q J0109+181             & 01:09:08.1 & +18:16:07  &  1   &  0.145   & 15.02 & HSP &   BL Lac           \\
NPMIG +41.0022          & 01:10:04.7 & +41:49:50  &  1   &  0.096   & 13.37 & LSP*&   BL.gal.domin     \\
1FGL J0110.0+6806       & 01:10:12.8 & +68:05:41  &  1   &  0.290   & 14.86 & ISP &   bl.un    \\
S4 0108+38              & 01:11:37.3 & +39:06:28  &  12  &  0.668   & 12.24 & LSP*&   bl.un    \\
S2 0109+22              & 01:12:05.8 & +22:44:38  &  1   &  0.265   & 14.33 & ISP &   BL Lac   \\
PKS 0118-272            & 01:20:31.6 & -27:01:24  &  3   &  0.560   & 14.33 & ISP &   BL Lac   \\
1ES 0120+340            & 01:23:08.5 & +34:20:47  &  1   &  0.272   & 17.46 & HSP &   BL Lac   \\
MS 0122.1+0903          & 01:24:44.5 & +09:18:49  &  3   &  0.338   & 13.75 & LSP*&   BL Lac   \\
PKS 0138-097            & 01:41:25.8 & -09:28:43  &  13  &  0.733   & 13.04 & LSP &   BL Lac   \\
PKS 0140-059            & 01:42:38.8 & -05:44:01  &  1   &  0       & 13.56 & LSP*&   BL Lac   \\
1ES 0145+138            & 01:48:29.7 & +14:02:18  &  8   &  0.125   & 14.27 & ISP &   BL.gal.domin    \\
8C 0149+710             & 01:53:25.8 & +71:15:06  &  2   &  0.022   & 15.69 & HSP &   BL.gal.domin    \\
87GB 0156.9+1032        & 01:59:34.4 & +10:47:07  &  6   &  0.195   & 16.03 & HSP &   BL Lac   \\
MS 0158.5+0019          & 02:01:06.1 & +00:34:00  &  4   &  0.298   & 16.82 & HSP*&   BL Lac   \\
PKS 0202+14             & 02:04:50.4 & +15:14:11  &  4   &  0.833   & 12.71 & LSP &   bl.un    \\
S5 0205+72              & 02:09:51.7 & +72:29:26  &  1   &  0.895   & 12.71 & LSP*&   bl.un    \\
Z 0214+083              & 02:17:17.0 & +08:37:03  &  7   &  1.400   & 13.79 & LSP &   BL Lac   \\
OD 330                  & 02:21:05.5 & +35:56:13  &  2   &  0.944   & 13.16 & LSP &   bl.un    \\
PKS 0219-164            & 02:22:00.7 & -16:15:16  &  1   &  0.698   & 14.58 & ISP &   FSRQ     \\
3C 66A                  & 02:22:39.6 & +43:02:07  &  5   &  0.444   & 15.09 & HSP &   BL Lac   \\
1ES 0229+200            & 02:32:48.6 & +20:17:17  &  5   &  0.140   & 15.48 & HSP &   BL.gal.domin     \\
AO 0235+164             & 02:38:38.8 & +16:36:59  &  28  &  0.940   & 13.31 & LSP &   BL Lac   \\
PKS 0245-167            & 02:48:07.7 & -16:31:46  &  3   &  0       & 12.84 & LSP*&   BL Lac   \\
RXS J0250.6+1712        & 02:50:38.0 & +17:12:08  &  7   &  1.100   & 16.41 & HSP &   bl.un    \\
PKS 0301-243            & 03:03:26.5 & -24:07:13  &  5   &  0.260   & 15.43 & HSP &   BL Lac   \\
4C 47.08                & 03:03:35.2 & +47:16:16  &  5   &  0.475   & 14.00 & ISP &   BL Lac   \\
PKS 0306+102            & 03:09:03.6 & +10:29:16  &  15  &  0.863   & 13.14 & LSP &   FSRQ     \\
RXS J0314.3+0620        & 03:14:23.9 & +06:19:57  &  5   &  0       & 15.92 & HSP*&   BL Lac   \\
RXS J0316.1+0904        & 03:16:12.9 & +09:04:43  &  6   &  0       & 15.77 & ISP &   BL Lac   \\
3C 84                   & 03:19:48.1 & +41:30:42  &  6   &  0.018   & 13.57 & LSP &   bl.un    \\
MS 0317.0+1834          & 03:19:51.8 & +18:45:35  &  6   &  0.190   & 16.99 & HSP &   BL.gal.domin     \\
2E 0323+0214            & 03:26:13.9 & +02:25:14  &  7   &  0.147   & 15.93 & HSP &   BL Lac   \\
PKS 0338-214            & 03:40:35.6 & -21:19:31  &  5   &  0.223   & 13.49 & LSP &   BL Lac   \\
PKS 0346-163            & 03:48:39.2 & -16:10:17  &  8   &  0       & 13.45 & LSP &   BL Lac   \\
S5 0346+800             & 03:54:46.1 & +80:09:28  &  1   &  0       & 12.60 & LSP &   cand.    \\
PKS 0357-264            & 03:59:33.6 & -26:15:31  &  3   &  1.470   & 13.16 & LSP &   BL Lac   \\
PKS 0406+121            & 04:09:22.1 & +12:17:39  &  19  &  1.020   & 13.16 & LSP*&   BL Lac   \\
2E 0414+0057            & 04:16:52.4 & +01:05:24  &  6   &  0.287   & 16.64 & HSP &   BL Lac   \\
MS 0419.3+1943          & 04:22:18.5 & +19:50:53  &  2   &  0.512   & 12.62 & LSP*&   BL Lac   \\
PKS 0420+022            & 04:22:52.2 & +02:19:27  &  9   &  2.277   & 12.35 & LSP*&   FSRQ     \\
PKS 0422+004            & 04:24:46.8 & +00:36:07  &  14  &  0.310   & 14.10 & ISP &   BL Lac   \\
MCG -01.12.005          & 04:25:51.3 & -08:33:39  &  4   &  0.039   & 12.80 & LSP*&   BL.gal.domin     \\
3C 120                  & 04:33:11.0 & +05:21:15  &  5   &  0.033   & 14.30 & ISP*&   bl.un    \\
2EG J0432+2910          & 04:33:37.8 & +29:05:55  &  2   &  0.970   & 13.40 & LSP &   BL Lac   \\
PKS 0439-299            & 04:41:19.5 & -29:52:35  &  2   &  0       & 13.91 & LSP*&   cand.    \\
PKS 0446+11             & 04:49:07.6 & +11:21:28  &  20  &  1.207   & 12.76 & LSP &   FSRQ     \\
\end{tabular}
\end{table}

\setcounter{table}{5}
\begin{table}
\caption{\label{tab:param2}Blazar sample and some parameters. Continued.}
\centering
\begin{tabular}{llllllll}\hline
Source & R.A. (J2000) & Dec (J2000) & N & z & $\log$ $\nu_{peak}^{S}$ & SED class & AGN class\\
\hline
PKS 0459+135            & 05:02:33.2 & +13:38:11  &  8   &  0       & 13.18 & LSP*&   BL Lac   \\
Q 0458+6530             & 05:03:05.8 & +65:34:01  &  1   &  0       & 16.45 & HSP*&   cand.    \\
RXS J0505.5+0416        & 05:05:34.7 & +04:15:54  &  6   &  0.027   & 15.96 & HSP &   BL Lac   \\
1ES 0502+675            & 05:07:56.1 & +67:37:24  &  6   &  0.416   & 18.20 & HSP*&   BL Lac   \\
S5 0454+84              & 05:08:42.3 & +84:32:04  &  3   &  1.340   & 13.28 & LSP*&   BL Lac   \\
MG 0509+0541            & 05:09:25.9 & +05:41:35  &  3   &  0       & 14.22 & ISP &   BL Lac   \\
4U 0506-03              & 05:09:39.0 & -04:00:36  &  14  &  0.304   & 17.13 & HSP*&   cand.    \\
1FGL J0515.2+7355       & 05:16:31.2 & +73:51:08  &  1   &  0.249   & 16.02 & HSP &   BL Lac   \\
PKS 0524+034            & 05:27:32.7 & +03:31:31  &  7   &  0       & 12.94 & LSP*&   BL Lac   \\
1WGA J0536.4-3342       & 05:36:29.1 & -33:43:02  &  1   &  0       & 16.05 & HSP*&   BL Lac   \\
HB89 0548-322           & 05:50:40.6 & -32:16:17  &  1   &  0.069   & 17.23 & HSP*&   BL.gal.domin     \\
B3 0609+413             & 06:12:51.1 & +41:22:37  &  2   &  0       & 14.49 & ISP &   BL Lac   \\
MS 0607.9+7108          & 06:13:43.3 & +71:07:26  &  1   &  0.267   & 15.00 & ISP*&   BL Lac   \\
J0617+5701              & 06:17:16.9 & +57:01:16  &  2   &  0       & 13.45 & LSP &   BL Lac   \\
87GB 06216+4441         & 06:25:18.2 & +44:40:01  &  7   &  0       & 14.55 & ISP &   BL Lac   \\
PKS 0627-199            & 06:29:23.7 & -19:59:19  &  1   &  0       & 13.23 & LSP &   BL Lac   \\
1ES 0647+250            & 06:50:46.5 & +25:03:00  &  2   &  0.203   & 16.42 & HSP &   BL Lac   \\
B3 0651+428             & 06:54:43.5 & +42:47:58  &  1   &  0.126   & 13.01 & LSP*&   BL.gal.domin     \\
4C +42.22               & 06:56:10.6 & +42:37:02  &  5   &  0.059   & 15.39 & HSP &   BL.gal.domin     \\
2MASS J06562263-2403194 & 06:56:22.6 & -24:03:19  &  1   &  0.371   & 15.30 & HSP*&   BL Lac   \\
J0707+6110              & 07:07:00.6 & +61:10:11  &  3   &  0       & 13.48 & LSP*&   BL Lac   \\
EXO 0706.1+5913         & 07:10:30.1 & +59:08:21  &  6   &  0.125   & 16.99 & HSP &   BL Lac   \\
B3 0707+476             & 07:10:46.1 & +47:32:11  &  2   &  1.292   & 13.24 & ISP &   BL Lac   \\
GB2 0716+332            & 07:19:19.4 & +33:07:09  &  1   &  0.779   & 14.08 & ISP &   FSRQ     \\
S5 0716+714             & 07:21:53.4 & +71:20:36  &  6   &  0.300   & 13.99 & LSP &   BL Lac   \\
PKS 0723-008            & 07:25:50.6 & -00:54:56  &  4   &  0.128   & 13.41 & LSP &   bl.un    \\
PKS 0735+17             & 07:38:07.4 & +17:42:19  &  15  &  0.424   & 13.98 & LSP &   BL Lac   \\
4C +54.15               & 07:53:01.3 & +53:52:59  &  3   &  0.200   & 13.28 & LSP &   BL Lac   \\
GB 0751+485             & 07:54:45.6 & +48:23:50  &  1   &  0.377   & 13.84 & LSP &   BL Lac   \\
PKS 0754+100            & 07:57:06.7 & +09:56:35  &  24  &  0.266   & 13.49 & LSP &   BL Lac   \\
PKS 0808+019            & 08:11:26.6 & +01:46:52  &  16  &  0.930   & 13.23 & LSP &   BL Lac   \\
1WGA J0816.0-0736       & 08:16:04.3 & -07:35:59  &  4   &  0.040   & 14.76 & ISP &   BL.gal.domin     \\
J0817-0933              & 08:17:49.7 & -09:33:30  &  2   &  0       & 14.05 & ISP &   BL Lac   \\
OJ 425                  & 08:18:16.0 & +42:22:45  &  5   &  0.530   & 12.99 & LSP &   BL Lac   \\
PKS 0818-128            & 08:20:57.4 & -12:58:59  &  6   &  0.074   & 14.72 & ISP &   BL Lac   \\
4C 22.21                & 08:23:24.7 & +22:23:03  &  4   &  0.951   & 13.27 & LSP*&   BL Lac   \\
PKS 0823+033            & 08:25:50.3 & +03:09:24  &  22  &  0.506   & 13.18 & LSP &   BL Lac   \\
PKS 0823-223            & 08:26:01.5 & -22:30:27  &  3   &  0.910   & 14.44 & ISP &   BL Lac   \\
PKS 0829+046            & 08:31:48.9 & +04:29:39  &  7   &  0.174   & 13.71 & LSP &   BL Lac   \\
1H 0827+089             & 08:31:55.1 & +08:47:43  &  5   &  0.941   & 13.15 & LSP*&   FSRQ     \\
OJ 448                  & 08:32:23.2 & +49:13:20  &  2   &  0.548   & 12.90 & LSP &   BL Lac   \\
TEX 0836+182            & 08:39:30.7 & +18:02:47  &  14  &  0.280   & 14.22 & ISP &   BL Lac   \\
PKS 0837+035            & 08:39:49.2 & +03:19:53  &  5   &  1.570   & 12.79 & LSP &   BL Lac   \\
RXS J0847.2+1133        & 08:47:12.9 & +11:33:52  &  4   &  0.198   & 15.95 & HSP &   BL Lac   \\
2MASS J08475674-0703169 & 08:47:56.7 & -07:03:17  &  1   &  0       & 13.53 & LSP &   BL Lac   \\
US1889                  & 08:54:09.8 & +44:08:30  &  3   &  0.382   & 14.16 & ISP &   BL Lac   \\
OJ 287                  & 08:54:48.8 & +20:06:30  &  11  &  0.306   & 13.73 & LSP &   BL Lac   \\
NPM1G -09.0307          & 09:08:02.2 & -09:59:37  &  7   &  0.054   & 12.91 & LSP*&   bl.un    \\
3C 216.0                & 09:09:33.4 & +42:53:46  &  7   &  0.670   & 13.31 & LSP*&   bl.un    \\
RXS J09130-2103         & 09:13:00.1 & -21:03:20  &  1   &  0.198   & 16.50 & HSP &   BL Lac   \\
B2 0912+29              & 09:15:52.3 & +29:33:24  &  3   &  0.101   & 15.59 & HSP &   BL Lac   \\
B2 0922+31B             & 09:25:43.6 & +31:27:10  &  1   &  0.260   & 13.27 & LSP &   BL Lac   \\
J09291544+5013360       & 09:29:15.4 & +50:13:35  &  4   &  0.370   & 14.14 & ISP &   BL Lac   \\
1ES 0927+500            & 09:30:37.5 & +49:50:25  &  5   &  0.188   & 15.29 & HSP &   BL Lac   \\
B2 0927+35              & 09:30:55.2 & +35:03:37  &  10  &  0       & 12.67 & LSP*&   BL Lac   \\
B2 0937+26              & 09:40:14.7 & +26:03:29  &  5   &  0.498   & 13.33 & LSP*&   BL Lac   \\
RXS J09449-1347         & 09:44:59.2 & -13:47:51  &  1   &  0       & 14.24 & ISP*&   cand.    \\
2FGL J0945.9+5751       & 09:45:42.2 & +57:57:47  &  1   &  0.229   & 14.73 & ISP &   BL Lac   \\
RXS J09530-0840         & 09:53:02.6 & -08:40:18  &  1   &  0       & 15.29 & HSP &   BL Lac   \\
S4 0954+65              & 09:58:47.2 & +65:33:54  &  6   &  0.367   & 13.49 & LSP &   BL Lac   \\
\end{tabular}
\end{table}

\setcounter{table}{5}
\begin{table}
\caption{\label{tab:param3}Blazar sample and some parameters. Continued.}
\centering
\begin{tabular}{llllllll}\hline
Source & R.A. (J2000) & Dec (J2000) & N & z & $\log$ $\nu_{peak}^{S}$ & SED class & AGN class\\
\hline
4C22.25                 & 10:00:21.9 & +22:33:18  &  7   &  0.419   & 14.39 & ISP*&   bl.un    \\
J1008+0621              & 10:08:00.8 & +06:21:21  &  2   &  0.650   & 14.21 & ISP*&   BL Lac   \\
RXS J1008.1+4705        & 10:08:11.3 & +47:05:20  &  3   &  0.343   & 14.49 & ISP &   BL Lac   \\
PKS 1008+013            & 10:11:15.6 & +01:06:42  &  1   &  1.275   & 13.01 & LSP*&   BL Lac   \\
NRAO 350                & 10:12:13.3 & +06:30:57  &  11  &  0.727   & 14.98 & ISP &   BL Lac   \\
RXS J1012.7+4229        & 10:12:44.3 & +42:29:57  &  8   &  0.364   & 16.81 & HSP &   BL Lac   \\
2FGL J1019.8+6322       & 10:19:50.8 & +63:20:01  &  1   &  2.025   & 13.24 & LSP &   BL Lac   \\
RXS J1022.7-0112        & 10:22:43.9 & -01:13:02  &  6   &  0       & 16.64 & HSP &   BL Lac   \\
SDSS J10326+6623        & 10:32:39.0 & +66:23:23  &  1   &  2.212   & 14.08 & ISP &   BL Lac   \\
J1036+1233              & 10:36:40.3 & +12:33:38  &  1   &  0       & 14.35 & ISP*&   BL Lac   \\
TEX 1040+244            & 10:43:09.0 & +24:08:35  &  15  &  0.560   & 12.77 & LSP &   FSRQ     \\
S5 1044+719             & 10:48:27.6 & +71:43:35  &  2   &  1.150   & 13.19 & LSP &   bl.un    \\
GB6 J1054+2210          & 10:54:30.6 & +22:10:54  &  1   &  1.539   & 14.60 & ISP &   BL Lac   \\
RX J10578-2753          & 10:57:50.7 & -27:54:11  &  1   &  0.092   & 15.64 & HSP &   BL Lac   \\
B3 1055+433             & 10:58:02.9 & +43:04:41  &  1   &  2.204   & 12.68 & LSP*&   BL Lac   \\
4C 01.28                & 10:58:29.6 & +01:33:58  &  8   &  0.890   & 13.18 & LSP &   bl.un    \\
MRK 421                 & 11:04:27.2 & +38:12:32  &  13  &  0.031   & 17.07 & HSP &   BL Lac   \\
RXS J1110.6+7133        & 11:10:37.5 & +71:33:56  &  1   &  0       & 15.55 & HSP &   BL Lac   \\
FIRST J1117.6+2548      & 11:17:40.4 & +25:48:46  &  1   &  0.360   & 15.60 & HSP &   BL Lac   \\
EXO 1118                & 11:20:48.0 & +42:12:12  &  1   &  0.124   & 17.18 & HSP &   BL Lac   \\
CGRaBS J1121-0711       & 11:21:42.1 & -07:11:06  &  1   &  0       & 12.44 & LSP*&   BL Lac   \\
J112402.70+23           & 11:24:02.7 & +23:36:45  &  5   &  1.549   & 13.12 & LSP &   FSRQ     \\
J1132+0034              & 11:32:45.6 & +00:34:27  &  4   &  1.223   & 14.07 & ISP &   BL Lac   \\
MS 1133.7+1618          & 11:36:17.6 & +16:01:53  &  3   &  0.574   & 12.16 & LSP*&   BL Lac   \\
MRK 180                 & 11:36:26.4 & +70:09:27  &  1   &  0.045   & 15.77 & HSP &   BL Lac   \\
A1137+1544              & 11:40:23.4 & +15:28:09  &  1   &  0.244   & 15.99 & HSP &   BL Lac   \\
GB6 B1144+3517          & 11:47:22.1 & +35:01:07  &  2   &  0.063   & 14.60 & ISP*&   BL.gal.domin    \\
J1148+1840              & 11:48:37.7 & +18:40:09  &  5   &  0.405   & 13.10 & LSP*&   BL Lac   \\
EXO 1149.9+2455         & 11:49:30.3 & +24:39:27  &  3   &  0.402   & 14.61 & ISP*&   BL Lac   \\
B2 1147+24              & 11:50:19.2 & +24:17:54  &  15  &  0.200   & 14.02 & ISP &   BL Lac   \\
RXS J1151.4+5859        & 11:51:24.6 & +58:59:17  &  1   &  0.118   & 14.72 & ISP &   BL Lac   \\
SBS 1200+608            & 12:03:03.5 & +60:31:19  &  1   &  0.065   & 14.93 & ISP &   BL Lac   \\
J1206+0529              & 12:06:58.0 & +05:29:52  &  4   &  0.791   & 13.85 & LSP*&   BL Lac   \\
CGRaBS J1209-2032       & 12:09:14.6 & -20:32:39  &  1   &  0.404   & 12.20 & LSP*&   BL Lac   \\
B3 1206+416             & 12:09:22.7 & +41:19:41  &  1   &  0.377   & 13.85 & LSP &   BL Lac   \\
1ES 1212+078            & 12:15:10.9 & +07:32:04  &  5   &  0.136   & 13.27 & LSP*&   BL.gal.domin    \\
GB6 B1215+3023          & 12:17:52.0 & +30:07:00  &  3   &  0.130   & 15.26 & HSP &   BL Lac   \\
PKS 1215-002            & 12:17:58.7 & -00:29:46  &  1   &  0.419   & 13.51 & LSP &   BL Lac   \\
GB2 1217+348            & 12:20:08.2 & +34:31:21  &  1   &  0.643   & 13.95 & LSP &   BL Lac   \\
PG 1218+304             & 12:21:21.9 & +30:10:37  &  5   &  0.182   & 16.66 & HSP &   BL Lac   \\
1WGA J1221.5+2813       & 12:21:31.6 & +28:13:58  &  6   &  0.102   & 14.44 & ISP &   BL Lac   \\
S5 1221+80              & 12:23:40.4 & +80:40:04  &  5   &  0       & 13.22 & LSP &   BL Lac   \\
RXS J12302+2517         & 12:30:14.0 & +25:18:06  &  6   &  0.135   & 14.91 & ISP &   BL Lac   \\
2E 1258+1437            & 12:31:23.9 & +14:21:25  &  2   &  0.260   & 15.03 & HSP &   BL Lac   \\
BZB J1235+1700          & 12:35:28.8 & +17:00:36  &  1   &  0.381   & 14.01 & ISP*&   BL.gal.domin    \\
FIRST J1236.3+3900      & 12:36:23.0 & +39:00:01  &  1   &  0.390   & 14.78 & ISP &   BL Lac   \\
RX J12370+3020          & 12:37:05.5 & +30:20:05  &  1   &  0.700   & 12.20 & LSP*&   BL Lac   \\
RXS J12416+3440         & 12:41:41.4 & +34:40:31  &  7   &  0       & 12.27 & LSP*&   BL Lac   \\
1ES 1239+069            & 12:41:48.3 & +06:36:01  &  3   &  0.150   & 15.06 & HSP &   BL Lac   \\
RX J12418-1455          & 12:41:49.3 & -14:55:58  &  1   &  0       & 15.88 & HSP &   BL Lac   \\
Ton 116                 & 12:43:12.7 & +36:27:43  &  3   &  1.065   & 16.15 & HSP &   BL Lac   \\
S5 1250+53              & 12:53:11.9 & +53:01:11  &  5   &  1.084   & 13.94 & LSP &   BL Lac   \\
PKS 1256-229            & 12:59:08.4 & -23:10:38  &  1   &  0.481   & 13.62 & LSP &   bl.un    \\
\end{tabular}
\end{table}

\setcounter{table}{5}
\begin{table}
\caption{\label{tab:param4}Blazar sample and some parameters. Continued.}
\centering
\begin{tabular}{llllllll}\hline
Source & R.A. (J2000) & Dec (J2000) & N & z & $\log$ $\nu_{peak}^{S}$ & SED class & AGN class\\
\hline
FIRST J1301.7+4056      & 13:01:45.6 & +40:56:24  &  1   &  0.649   & 13.04 & LSP*&   BL Lac   \\
GB6 B1300+5804          & 13:02:52.4 & +57:48:37  &  2   &  1.088   & 12.28 & LSP &   bl.un    \\
RXS J1302.9+5056        & 13:02:55.5 & +50:56:17  &  1   &  0.688   & 12.42 & LSP*&   BL Lac   \\
MC2 1307+12             & 13:09:33.9 & +11:54:24  &  14  &  0.318   & 13.32 & LSP &   BL Lac   \\
OP 313                  & 13:10:28.6 & +32:20:43  &  10  &  0.997   & 13.30 & LSP &   bl.un    \\
PKS 1309-216            & 13:12:31.5 & -21:56:24  &  1   &  1.491   & 15.46 & HSP &   BL Lac   \\
RXS J1319.5+1405        & 13:19:31.7 & +14:05:34  &  5   &  0.572   & 15.49 & HSP &   BL Lac   \\
J132247.40+3216         & 13:22:47.3 & +32:16:08  &  5   &  0       & 15.13 & HSP*&   BL Lac   \\
RXS J1326.2+1230        & 13:26:17.6 & +12:29:58  &  5   &  0.204   & 12.33 & LSP*&   BL.gal.domin     \\
J132952.86+3154         & 13:29:52.8 & +31:54:11  &  1   &  0       & 12.44 & LSP*&   cand.    \\
SDSS J13338+5057        & 13:33:53.7 & +50:57:36  &  1   &  1.362   & 14.32 & ISP &   bl.un    \\
MS1332.6-2935           & 13:35:29.7 & -29:50:39  &  1   &  0.513   & 14.86 & ISP &   BL Lac   \\
RXS J1341.0+3959        & 13:41:05.0 & +39:59:45  &  6   &  0.172   & 14.63 & ISP &   BL.gal.domin    \\
J134916-141316          & 13:49:16.0 & -14:13:17  &  1   &  0.253   & 13.20 & LSP*&   BL Lac   \\
PKS1350+148             & 13:53:22.8 & +14:35:39  &  2   &  0.807   & 13.57 & LSP &   BL Lac   \\
RXS J1353.4+5601        & 13:53:28.0 & +56:00:56  &  2   &  0.404   & 12.18 & LSP*&   BL Lac   \\
MC 1400+162             & 14:02:44.5 & +15:59:57  &  12  &  0.244   & 14.44 & ISP*&   BL Lac   \\
MS 1402.3+0416          & 14:04:51.0 & +04:02:02  &  5   &  0.344   & 15.83 & HSP &   BL Lac   \\
MS 1407.9+5954          & 14:09:23.4 & +59:39:40  &  1   &  0.496   & 15.04 & HSP*&   BL Lac   \\
PKS 1407+022            & 14:10:04.6 & +02:03:06  &  10  &  0       & 13.24 & LSP*&   BL Lac   \\
RXS J1410.5+6100        & 14:10:31.7 & +61:00:10  &  5   &  0.384   & 14.93 & ISP*&   BL Lac   \\
PKS 1413+135            & 14:15:58.8 & +13:20:24  &  11  &  0.247   & 12.96 & LSP &   bl.un    \\
2E 1415+2557            & 14:17:56.6 & +25:43:25  &  2   &  0.237   & 15.45 & HSP &   BL Lac   \\
OQ 530                  & 14:19:46.6 & +54:23:14  &  4   &  0.153   & 13.98 & LSP &   BL Lac   \\
SDSS J14202+0614        & 14:20:13.6 & +06:14:28  &  1   &  0.625   & 13.49 & LSP*&   BL Lac   \\
PKS 1424+240            & 14:27:00.5 & +23:48:00  &  11  &  0.160   & 15.34 & HSP &   BL Lac   \\
RXS J1436.9+5639        & 14:36:57.7 & +56:39:24  &  1   &  0.150   & 16.88 & HSP &   BL Lac   \\
PKS 1437-153            & 14:39:56.8 & -15:31:50  &  7   &  0.636   & 13.46 & LSP &   BL Lac   \\
1ES 1440+122            & 14:42:48.3 & +12:00:40  &  6   &  0.162   & 16.35 & HSP &   BL Lac   \\
RXS J14495+2746         & 14:49:32.6 & +27:46:21  &  1   &  0.227   & 12.40 & LSP*&   BL.gal.domin     \\
B2 1451+26              & 14:53:53.5 & +26:48:33  &  3   &  0       & 13.12 & LSP*&   bl.un    \\
RXS J1456.0+5048        & 14:56:03.7 & +50:48:25  &  5   &  0.480   & 12.52 & LSP*&   BL Lac   \\
RXS J1458.4+4832        & 14:58:27.0 & +48:32:46  &  2   &  0.541   & 15.44 & HSP*&   BL Lac   \\
B3 1456+375             & 14:58:44.8 & +37:20:22  &  8   &  0.333   & 13.31 & LSP &   BL Lac   \\
TXS 1459+480            & 15:00:48.6 & +47:51:15  &  1   &  1.059   & 13.15 & LSP &   FSRQ     \\
PKS 1514+197            & 15:16:56.8 & +19:32:12  &  19  &  0.650   & 12.97 & LSP &   BL Lac   \\
PKS 1514-24             & 15:17:41.8 & -24:22:19  &  4   &  0.048   & 14.19 & ISP &   BL Lac   \\
PKS 1519-273            & 15:22:37.6 & -27:30:10  &  5   &  1.294   & 12.79 & LSP &   BL Lac   \\
1ES 1533+535            & 15:35:00.8 & +53:20:37  &  5   &  0.890   & 16.99 & HSP &   BL Lac   \\
MS 1534.2+0148          & 15:36:46.8 & +01:37:59  &  5   &  0.311   & 14.64 & ISP*&   BL Lac   \\
4C 14.60                & 15:40:49.4 & +14:47:45  &  20  &  0.605   & 13.53 & LSP &   BL Lac   \\
RXS J1542.9+6129        & 15:42:56.9 & +61:29:55  &  2   &  0.117   & 14.64 & ISP &   BL Lac   \\
RXS J1544.3+0458        & 15:44:18.7 & +04:58:22  &  6   &  0.326   & 12.80 & LSP*&   BL.gal.domin     \\
PG 1553+11              & 15:55:43.1 & +11:11:24  &  10  &  0.326   & 14.12 & ISP &   BL Lac   \\
MYC 1557+566            & 15:58:48.2 & +56:25:14  &  1   &  0.300   & 14.21 & ISP &   BL Lac   \\
CGRaBS J1603+1105       & 16:03:41.9 & +11:05:48  &  7   &  0.143   & 13.46 & LSP &   BL Lac   \\
PKS 1604+159            & 16:07:06.4 & +15:51:34  &  10  &  0.357   & 13.36 & LSP &   BL Lac   \\
RXS J1610.0+6710        & 16:10:04.1 & +67:10:26  &  1   &  0.067   & 16.20 & HSP*&   BL Lac   \\
SDSS J16183+3632        & 16:18:23.5 & +36:32:01  &  3   &  0.730   & 13.43 & LSP*&   BL Lac   \\
NGC 6251                & 16:32:31.9 & +82:32:16  &  3   &  0.025   & 13.73 & LSP &   bl.un    \\
CGRaBS J1642-0621       & 16:42:02.1 & -06:21:23  &  4   &  1.514   & 13.23 & LSP &   BL Lac   \\
1FGL J1647.4+4948       & 16:47:34.9 & +49:50:00  &  1   &  0.047   & 13.70 & LSP &   bl.un    \\
PKS 1648+015            & 16:51:03.6 & +01:29:23  &  4   &  0.400   & 12.88 & LSP*&   BL Lac   \\
MRK 501                 & 16:53:52.2 & +39:45:36  &  12  &  0.033   & 16.12 & HSP &   BL Lac   \\
SDSS J16581+6150        & 16:58:08.3 & +61:50:02  &  1   &  0.374   & 14.62 & ISP &   BL Lac   \\
\end{tabular}
\end{table}

\setcounter{table}{5}
\begin{table}
\caption{\label{tab:param5}Blazar sample and some parameters. Continued.}
\centering
\begin{tabular}{llllllll}\hline
Source & R.A. (J2000) & Dec (J2000) & N & z & $\log$ $\nu_{peak}^{S}$ & SED class & AGN class\\
\hline
PKS 1707-038            & 17:10:17.2 & -03:55:50  &  9   &  1.920   & 12.78 & LSP*&   FSRQ     \\
PGC 59947               & 17:15:22.9 & +57:24:40  &  1   &  0.027   & -     & -   &   BL.gal.domin     \\
TEX 1714-336            & 17:17:36.0 & -33:42:08  &  3   &  0       & 13.08 & LSP*&   cand.    \\
PKS 1717+177            & 17:19:13.1 & +17:45:06  &  11  &  0.137   & 13.33 & LSP*&   BL Lac   \\
H 1722+119              & 17:25:04.4 & +11:52:16  &  6   &  0.018   & 16.01 & HSP &   BL Lac   \\
BZB J1733+4519          & 17:33:28.8 & +45:19:50  &  1   &  0.317   & -     & -   &   BL.gal.domin     \\
OT 465                  & 17:39:57.1 & +47:37:58  &  1   &  0.950   & 13.19 & LSP &   BL Lac   \\
NPM1G +19.0510          & 17:43:57.9 & +19:35:09  &  7   &  0.083   & 15.76 & HSP &   BL.gal.domin     \\
J1745-0753              & 17:45:27.1 & -07:53:03  &  7   &  0       & 13.36 & LSP &   BL Lac   \\
B2 1743+39C             & 17:45:37.6 & +39:51:31  &  3   &  0.267   & 15.15 & HSP &   BL.gal.domin     \\
S4 1749+70              & 17:48:32.8 & +70:05:50  &  3   &  0.770   & 13.78 & LSP &   BL Lac   \\
PKS 1749+096            & 17:51:32.7 & +09:39:01  &  17  &  0.322   & 12.41 & LSP &   BL Lac   \\
RXS J1756.2+5522        & 17:56:15.9 & +55:22:18  &  5   &  0.407   & 16.46 & HSP &   BL Lac   \\
S5 1803+78              & 18:00:45.6 & +78:28:04  &  5   &  0.680   & 13.55 & LSP &   BL Lac   \\
3C 371                  & 18:06:50.6 & +69:49:28  &  3   &  0.046   & 14.69 & ISP &   BL Lac   \\
4C 56.27                & 18:24:07.0 & +56:51:01  &  3   &  0.664   & 13.19 & LSP &   BL Lac   \\
3C 380.0                & 18:29:31.7 & +48:44:46  &  3   &  0.695   & 13.27 & LSP &   bl.un    \\
1H 1914-194             & 19:17:44.8 & -19:21:30  &  1   &  0.137   & 15.41 & HSP &   BL Lac   \\
S4 1926+61              & 19:27:30.4 & +61:17:31  &  1   &  0       & 13.36 & LSP &   BL Lac   \\
S5 2007+77              & 20:05:30.9 & +77:52:43  &  6   &  0.342   & 13.53 & LSP &   BL Lac   \\
4C +72.28               & 20:09:52.3 & +72:29:19  &  2   &  0       & 12.90 & LSP &   BL Lac   \\
PKS 2012-017            & 20:15:15.1 & -01:37:33  &  6   &  0.520   & 14.43 & ISP &   BL Lac   \\
S5 2023+76              & 20:22:35.5 & +76:11:26  &  5   &  0.594   & 14.11 & ISP &   BL Lac   \\
PKS 2029+121            & 20:31:54.9 & +12:19:41  &  2   &  1.215   & 12.95 & LSP &   bl.un    \\
PKS 2032+107            & 20:35:22.3 & +10:56:06  &  6   &  0.601   & 14.03 & ISP &   FSRQ     \\
1ES 2037+521            & 20:39:23.5 & +52:19:49  &  1   &  0.053   & 16.47 & HSP &   cand.    \\
PKS 2047+039            & 20:50:06.2 & +04:07:49  &  12  &  0       & 13.08 & LSP &   BL Lac   \\
S5 2051+74              & 20:51:33.7 & +74:41:40  &  1   &  0       & 12.76 & LSP*&   cand.    \\
GB6 J2116+3339          & 21:16:14.5 & +33:39:20  &  1   &  0.350   & 15.33 & HSP &   BL Lac   \\
PKS 2131-021            & 21:34:10.2 & -01:53:17  &  14  &  0.557   & 13.19 & LSP &   BL Lac   \\
MS 2143.4+0704          & 21:45:52.3 & +07:19:27  &  5   &  0.237   & 14.58 & ISP &   BL Lac   \\
PKS 2149+17             & 21:52:24.8 & +17:34:37  &  15  &  0.871   & 13.38 & LSP &   BL Lac   \\
PKS 2155-304            & 21:58:52.0 & -30:13:32  &  3   &  0.116   & 15.97 & HSP &   BL Lac   \\
BL LAC                  & 22:02:43.2 & +42:16:40  &  19  &  0.069   & 13.39 & LSP &   BL Lac   \\
B2 2214+24B             & 22:17:00.8 & +24:21:46  &  5   &  0.505   & 13.37 & LSP &   BL Lac   \\
J221944+212056          & 22:19:44.1 & +21:20:53  &  2   &  0.200   & 13.88 & LSP*&   BL Lac   \\
PKS 2221-116            & 22:24:07.9 & -11:26:21  &  2   &  0.115   & 13.41 & LSP &   BL Lac   \\
PKS 2223-114            & 22:25:43.7 & -11:13:41  &  5   &  0.997   & 13.04 & LSP*&   BL Lac   \\
3C 446                  & 22:25:45.1 & -04:56:34  &  14  &  1.404   & 13.32 & LSP &   FSRQ     \\
FIRST J22279+0037       & 22:27:58.1 & +00:37:05  &  1   &  0       & 14.41 & ISP &   BL Lac   \\
RXS J2233.0+1335        & 22:33:01.1 & +13:36:02  &  5   &  0.214   & 14.67 & ISP*&   BL Lac   \\
PKS 2233-173            & 22:36:09.5 & -17:06:21  &  1   &  0.647   & 13.57 & LSP &   BL Lac   \\
PKS 2233-148            & 22:36:34.0 & -14:33:22  &  2   &  0.325   & 13.08 & LSP &   BL Lac   \\
B3 2238+410             & 22:41:07.2 & +41:20:11  &  1   &  0.726   & 13.32 & LSP &   BL Lac   \\
PKS 2240-260            & 22:43:26.4 & -25:44:30  &  3   &  0.774   & 13.22 & LSP &   BL Lac   \\
RGB J2243+203           & 22:43:54.7 & +20:21:03  &  1   &  0       & 15.58 & HSP &   BL Lac   \\
B3 2247+381             & 22:50:05.7 & +38:24:37  &  1   &  0.119   & 15.93 & HSP*&   BL Lac   \\
PKS 2251+006            & 22:54:04.4 & +00:54:20  &  2   &  0       & 12.92 & LSP*&   BL Lac   \\
PKS 2254+074            & 22:57:17.3 & +07:43:12  &  7   &  0.190   & 13.76 & LSP*&   BL Lac   \\
\end{tabular}
\end{table}

\setcounter{table}{5}
\begin{table}
\caption{\label{tab:param6}Blazar sample and some parameters. Continued.}
\centering
\begin{tabular}{llllllll}\hline
Source & R.A. (J2000) & Dec (J2000) & N & z & $\log$ $\nu_{peak}^{S}$ & SED class & AGN class\\
\hline
RXS J2304.6+3705        & 23:04:36.6 & +37:05:08  &  7   &  0       & 15.88 & HSP &   BL Lac   \\
PGC 1465934             & 23:13:57.3 & +14:44:23  &  1   &  0.163   & 15.42 & HSP &   BL.gal.domin     \\
Q J2319+161             & 23:19:43.4 & +16:11:50  &  2   &  0       & 12.63 & LSP*&   BL Lac   \\
GB6 J2325+3957          & 23:25:17.8 & +39:57:37  &  1   &  0       & 13.71 & LSP &   BL Lac   \\
1ES 2326+174            & 23:29:03.3 & +17:43:30  &  5   &  0.213   & 15.62 & HSP*&   BL Lac   \\
Q J2338+212             & 23:38:56.4 & +21:24:41  &  5   &  0.291   & 14.83 & ISP &   cand.    \\
MS 2336.5+0517          & 23:39:07.0 & +05:34:36  &  2   &  0.740   & 15.95 & HSP*&   BL Lac   \\
B2 2337+26              & 23:40:29.0 & +26:41:56  &  6   &  0.372   & 12.42 & LSP*&   cand.    \\
1FGL J2341.6+8015       & 23:40:53.7 & +80:15:13  &  1   &  0.274   & 15.64 & HSP &   BL Lac   \\
MS 2342.7-1531          & 23:45:22.3 & -15:15:06  &  1   &  0       & 16.18 & HSP*&   FSRQ     \\
MS J23492105+0534       & 23:49:21.0 & +05:34:40  &  4   &  0.419   & 13.58 & LSP*&   FSRQ     \\
MS 2347.4+1924          & 23:50:01.7 & +19:41:52  &  4   &  0.515   & 13.37 & LSP*&   bl.un    \\
RXS J2350.3-059         & 23:50:18.0 & -05:59:27  &  1   &  0.515   & 13.89 & LSP*&   cand.    \\
PKS 2354-02             & 23:57:25.1 & -01:52:15  &  7   &  0.812   & 13.38 & LSP &   BL Lac   \\
\hline
\multicolumn{8}{l}{}\\
\multicolumn{8}{l}{Col.~1~-- source name, Col.~2--3~-- R.A. and Dec (J2000.0), Col.~4~-- number of observing epochs at RATAN, Col.~5~-- redshift $z$ (BZCAT or Simbad),}\\
\multicolumn{8}{l}{Col.~6~-- logarithm of the synchrotron peak frequency $\nu_{peak}^{S}$ in Hz, Col.~7~-- SED class from BZCAT or \citet{2006A&A...445..441N},  we noted "*" blazars, for which the values of
$\log\nu_{peak}^{S}$  were calculated using ASDC SED builder,}\\
\multicolumn{8}{l}{Col.~8~-- AGN class from BZCAT}
\end{tabular}
\end{table}
\label{lastpage}
\end{document}